\documentclass[aps,prl,showpacs,twocolumn,groupedaddress,floatfix]{revtex4}
\usepackage{graphicx}
\usepackage{color}
\begin{document}

\title{Efficient spin transitions in inelastic electron tunneling spectroscopy}
\author{Nicol\'as Lorente$^{1}$}
\author{Jean-Pierre Gauyacq$^{2,3}$}
 \affiliation{$^{1}$ Centre d'Investigaci\'o en Nanoci\`encia i Nanotecnologia
 (CSIC-ICN), Campus de la UAB, Bellaterra, Spain\\
$^2$ CNRS, Laboratoire des Collisions Atomiques et Mol\'eculaires, UMR  8625, B\^atiment 351, Universit\'e Paris-Sud, 91405 Orsay CEDEX, 
France\\ 
$^3$ Universit\'e Paris-Sud, Laboratoire des Collisions Atomiques et Mol\'eculaires, UMR  8625, B\^atiment 351, Universit\'e Paris-Sud, 91405 Orsay CEDEX, 
France 
}
\date{\today}
\begin{abstract}
The excitation of the spin degrees of freedom of an adsorbed atom by tunneling 
electrons is computed using a strong coupling theory. The excitation process 
is shown to be a sudden switch between the initial state determined by the 
environmental anisotropy to an intermediate state given by the coupling to 
the tunnelling electron. This explains the observed large inelastic currents. 
Application is presented for Fe and Mn adsorbates on CuN monolayers on Cu(100).
First-principles calculations show the dominance of one collisional channel, 
leading to a quantitative agreement with the experiment.
\end{abstract}
\pacs{68.37.Ef, 72.10.-d, 73.23.-b, 72.25.-b}

\maketitle

The way electrons flow through atomic contacts has important
fundamental and technological implications~\cite{Agrait}. 
Electronic transport is a quantal process in which charge,
spin and vibrational degrees of
freedom are entangled leading to problems of intrinsic fundamental
interest. 
Technologically, the quest for minutarization is pushing the limits of devices
to the atomic scale, where the above transport properties will determine
the actual device functionalities.
An important issue is the appearance of inelastic effects where 
energy is taken from the electron flow into the different degrees 
of freedom of the system. Inelasticities lead to new regimes of transport
that contain relevant information on the atomic contact and have
been thus used to develop single atom and molecule 
spectroscopies~\cite{Stipe,Qiu,Heinrich}.

Inelastic electron tunneling spectroscopy (IETS) where electrons excite
vibrations leading to conductance steps at certain voltage thresholds~\cite{Stipe} has been extensively studied in the last 
years~\cite{Ho,Komeda,Bocquet,Morgenstern,Okabayashi}.
The inelastic change in conductance
 is within a few percent of the elastic conductance, mainly due
the smallness of the electron-vibration coupling~\cite{Paulsson,Monturet}. 
Recently,
Heinrich and co-workers have been able to develop a spin-resolved
spectroscopy using an STM~\cite{Heinrich,Hirjibehedin06,Hirjibehedin07,Otte}.
In magnetic  IETS~\cite{Heinrich}, the tunneling electron yields energy
to the spin of an adsorbed magnetic atom and in this way changes its orientation
by overcoming the magnetic anisotropy barrier of the atom on the surface.
 Magnetic transitions in the meV range could be observed in adsorbates partly decoupled from a metal 
substrate~\cite{Heinrich,Hirjibehedin06,Hirjibehedin07,Otte,Pc}.
As in vibrational IETS, the conductance presents a step at the energy threshold
however  
the changes in conductance at inelastic threshold can reach several 
hundreds percent.  This is at odds with previous 
treaments~\cite{Hirjibehedin07,Mats,Fernandez}
where first-order perturbation theory is used. 

In this letter, we present an all-order theory of the spin transitions 
IETS and apply it to the cases of Fe and  Mn 
adsorbates on a CuN monolayer on Cu, 
experimentally studied in  Refs.~\cite{Hirjibehedin07,Otte}. 
We compute the relative weights 
of both elastic and inelastic channels, leading to a quantitative account 
of the inelastic currents in the experimental observations. 
The theory reveals the nature of the inelastic transitions and explains 
the extremely large inelastic currents in these magnetic systems.

The general idea of our approach is the following. The spin of the adsorbate is 
in an initial state given by the anisotropy imposed by its environment 
and  by an external magnetic field, B. 
During the very short collisional time between the adsorbate and the 
tunneling electron, the electron spin couples with the adsorbate spin,
forming a transient collisional intermediate, whereas the interaction 
with the adsorbate environment can be neglected. This sudden switch between 
different coupling schemes of the adsorbate induces efficient transitions 
among magnetic states.
This excitation mechanism is not only found
in STM-induced spin flip.
Similar excitation processes have been shown to be very efficient for 
spin-forbidden electronic excitations  in 
electron-molecule collisions~\cite{O2} or in surface 
processes~\cite{Bahrim}, as well as for rotational IETS~\cite{RotationO2}.
	
\begin{figure}
\includegraphics[width=0.35\textwidth]{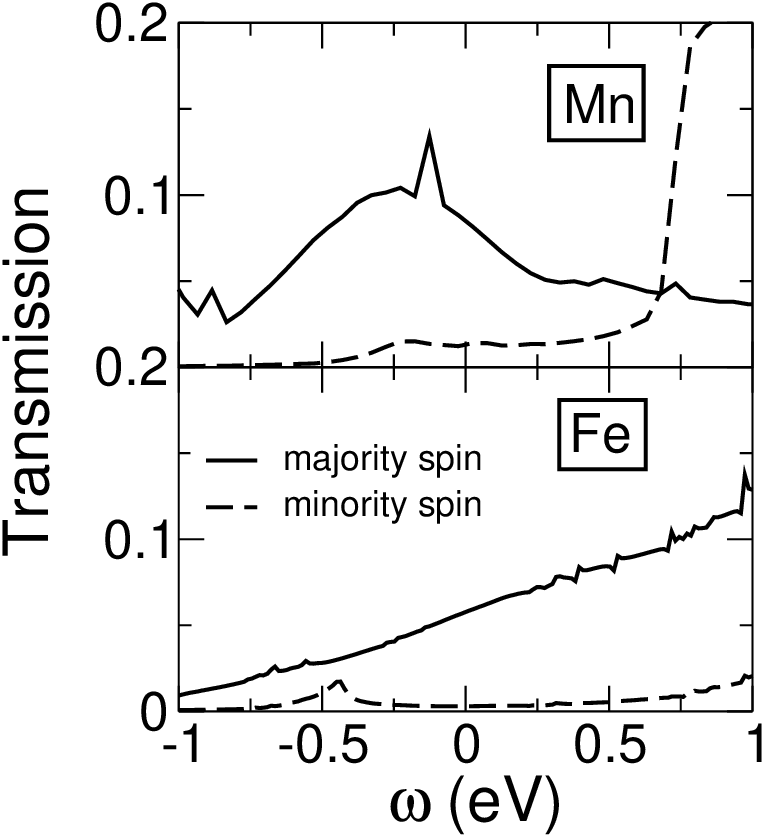}
\caption{Electron transmission as a function
of electron energy, $\omega$, in a tunneling junction described
by an atomic apex on a semi-infinite Cu(100) surface
 for the STM tip and  a Fe (lower pannel) a Mn (upper pannel) atom on a CuN
monolayer on a semi-infinite Cu(100) electrode. 
The magnetic atom--apex distance is 5.2 \AA~in the
present calculation. The full line is the transmission for the majority spin
 and the
dashed line for the minority one.
\label{figure1}}
\end{figure}

The energy losses associated with the magnetic anisotropy in the presence of 
a magnetic field, B,  have been modelled very efficiently in these 
systems~\cite{Hirjibehedin07,Otte} using the following Hamiltonian :	 				 
\begin{equation}
	H = g \mu_{B} \vec{B} \cdot \vec{S}  + D S_{z}^{2} + E (S_{x}^{2}-S_{y}^{2})
\label{hamiltonien}
\end{equation}
		 				
Where E and D are two constants describing the effect of the environment on 
the spin direction, g is the gyromagnetic factor  and  $\mu_{B}$ the Bohr 
magneton~\cite{Hirjibehedin07,Otte}. $\vec{S}$ is the spin operator of
the adsorbate and  $S_{x,y,z}$ its projections on the Cartesian axes. 
Diagonalisation of Hamiltonian~(\ref{hamiltonien}) yields the various possible 
$\phi_{n}$ states of the adsorbate spin in the system:		 				
\begin{equation}
\left|\phi_{n}\right\rangle = \sum_{M} C_{n,M} \left|S,M\right\rangle
\label{phi_n}
\end{equation}		 				
where $\left|S,M\right\rangle$ are eigenvectors of the $\vec{S}^2$, $S_z$  
operators. An electron injected from the STM tip  collides with the adsorbate 
on the surface and can cause inelastic transitions between the $\phi_{n}$  
states, which are recorded in an IETS experiment. In the present work, we use 
the modelling performed in Ref.~\cite{Hirjibehedin07,Otte} 
(D and E parameters, g and spin of the adsorbate), which very precisely 
reproduces the energy positions of the inelastic thresholds.

	In a first step, we compute the electron transmission through Fe 
and Mn adatoms on a CuN monolayer on Cu(100) by Density Functional Theory (DFT) with the {\tt Transiesta} code~\cite{Brandbyge}. 
We used the generalized gradient approximation (PBE) and a double-zeta local basis set where the contact region is modeled by a 7-atom slab, a CuN layer and
a Fe (Mn) atom, a vacuum gap of 5.2~\AA~ and a 5-atom slab  with an extra atom
for the tip region of the contact. The contact is relaxed
using the {\tt Siesta} method~\cite{Siesta}. Atomic forces are relaxed below 0.04 eV/\AA . The transmission is then computed for zero bias voltage, using the 
bulk Cu unit cell along the [100] direction as the primary unit of the two 
semi-infinite electrodes coupled to the contact region.  
On the energy scale relevant for the present magnetic IETS context, 
the majority spin transmission is 20 times the
minority spin one for the Fe junction, see Fig.~\ref{figure1}. 
Despite the fact that the actual spin state of the atom cannot be taken 
into account by DFT, these simulations can yield quantitative data
in spin transport~\cite{guo}.
Indeed, this difference implies that the transmission of minority-spin channels can 
be considered as suppressed. 
In addition, the transmission is seen to be almost flat as a function of 
electron energy, so that simple branching ratios can be used to obtain the relative value of the elastic and inelastic conductance of the system as a function of the STM bias.
For the Mn adsorbate, we find a similar result, though with a weaker dominance (factor 5) of the majority channel (Fig.~\ref{figure1}).

The branching ratios between elastic and inelastic conductance are determined 
making use of the following facts:

i) the rotation of the adsorbate spin, $\vec{S}$, due to the magnetic anisotropy, Eq.~(\ref{hamiltonien}), is slow compared to the electron-atom collision time so that we can use a sudden approximation, neglecting the effect of  Hamiltonian (\ref{hamiltonien}) during the collision.

ii) the spin of the tunnelling electron couples to the spin of the atom to define collision channels of total spin $S_T = S+1/2$ and $S-1/2$ that are linked to the asymptotic channels of the collision via:		 		
\begin{equation}
\left|S_{T},M_{T}\right\rangle = \sum_{m} CG_{S_{T},M_{T},m} \left|S,M=M_{T}-m \right\rangle 	\left|1/2,m\right\rangle
\label{intermediaire}
\end{equation}	 		
where the kets on the rhs correspond to the decoupled spins of the atom and of the tunnelling electron. $m$ is the projection of the electron spin on the $z$-axis. The $CG$ are Clebsch-Gordan coefficients that give the weight of the various $\left|S,M\right\rangle$ states in the collision channels. From Eqs.~(\ref{phi_n})
 and (\ref{intermediaire}), we can express the collision channel states as functions of the 
initial and final states of the collision: 	
\begin{equation}
	\left|j\right\rangle  =\left|S_{T},M_{T}\right\rangle =
	\sum_{n,m} A_{j,n,m} \left|\phi_{n}\right\rangle 	\left|1/2,m\right\rangle 
\label{Ajnm}
\end{equation}			
It yields the weight of the various anisotropy states in the collision channels associated to the  total spin, $S_T$.


iii) From the DFT result, we only consider the maximum spin intermediate state 
($S_T = 5/2$) for the Fe adsorbate and ($S_T = 3$) for Mn adsorbate.

iv) From Eqs.~(\ref{Ajnm}), we can derive the amplitude for transitions from $\left|\phi_{n}\right\rangle  	\left|1/2,m\right\rangle$ to $\left|\phi_{n'}\right\rangle  	\left|1/2,m'\right\rangle$ through the intermediate $j$ as proportional to 
 the product $A_{j,n,m} A_{j,n',m'}$.
The contributions from the different intermediate states are then added coherently
for the indistinguishable channels (same final $(n',m')$ state  for a given $(n,m)$ 
initial state) and incoherently for the distinguishable channels leading to the 
relative excitation probability (branching ratio) of the different excited states:	 					
	
\begin{equation}
	W_{n\rightarrow n'} =
	 \frac{\sum\limits_{m,m'} \left|\sum\limits_{j} A_{j,n,m} A_{j,n',m'}\right|^{2}  }  { \sum\limits_{n',m,m'} \left|\sum\limits_{j} A_{j,n,m} A_{j,n',m'}\right|^{2} }
\label{fraction}
\end{equation}

Note that because of the dominance of one $S_T$ intermediate state in the conductance, the sum over $j$ only concerns the $M$ sublevels, i.e.  the orientation of the  spin of the intermediate state; the corresponding contributions only differ by spin coupling coefficients and add coherently.
The sum over $j$  runs over the	$S_T = 5/2$ (resp. $S_T = 3$) intermediates for the Fe 
(respectively Mn) adsorbates, and the sum over $m$ and $m'$
 	concerns the spin up and down of the collisional electron.	
Equation ~(\ref{fraction}) above has been derived for an unpolarised incident electron; it can be easily generalized to yield spin-resolved transitions.
		 					
\begin{figure}
\includegraphics[width=0.35\textwidth]{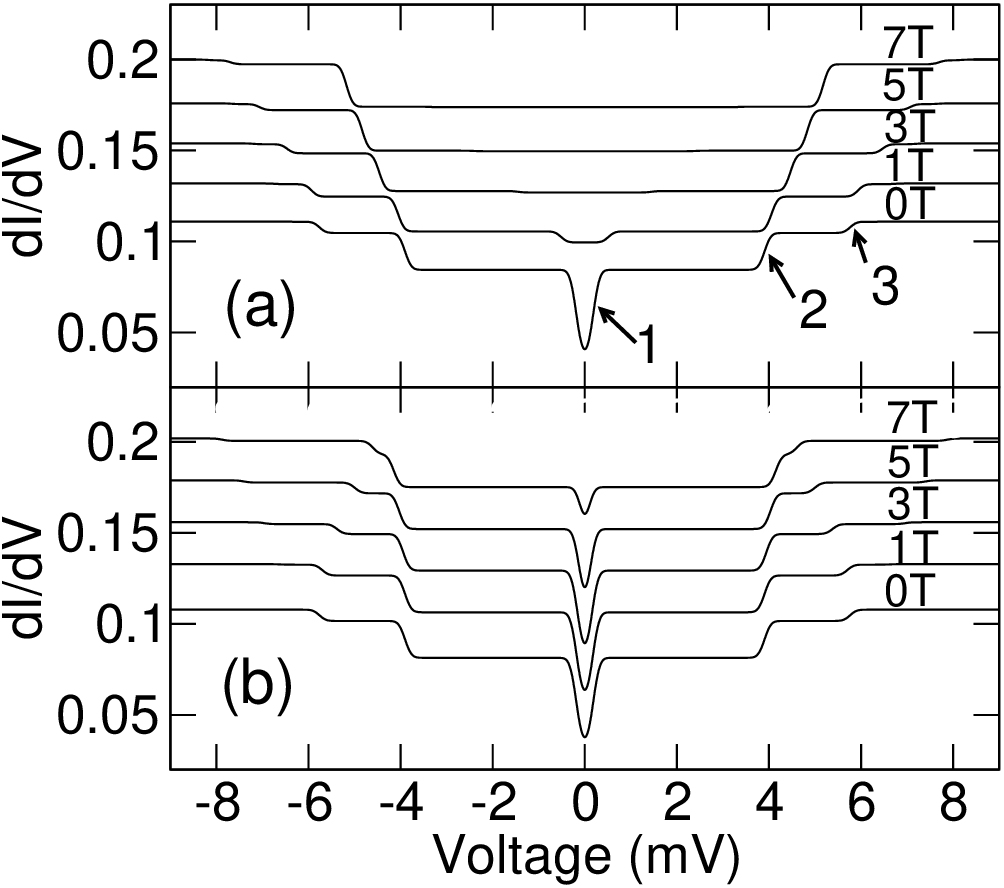}
\caption{
Computed conductance  for a Fe atom on  a CuN
monolayer on Cu(100) in atomic units. 
The conductances for increasing magnetic field B = 0
are  vertically displaced for representation purposes. The  B field is oriented along the N axis in part (a) and along the hollow axis on the surface in part (b). 
\label{figure2}}
\end{figure}

Equation ~(\ref{fraction}) is the basis of the present work. It yields the relative weight of the elastic and inelastic channels in the conductance. This expression is a direct consequence of spin coupling  and magnetic anisotropy, associated to the dominance of the majority spin conductance.

\begin{figure}
\includegraphics[width=0.35\textwidth]{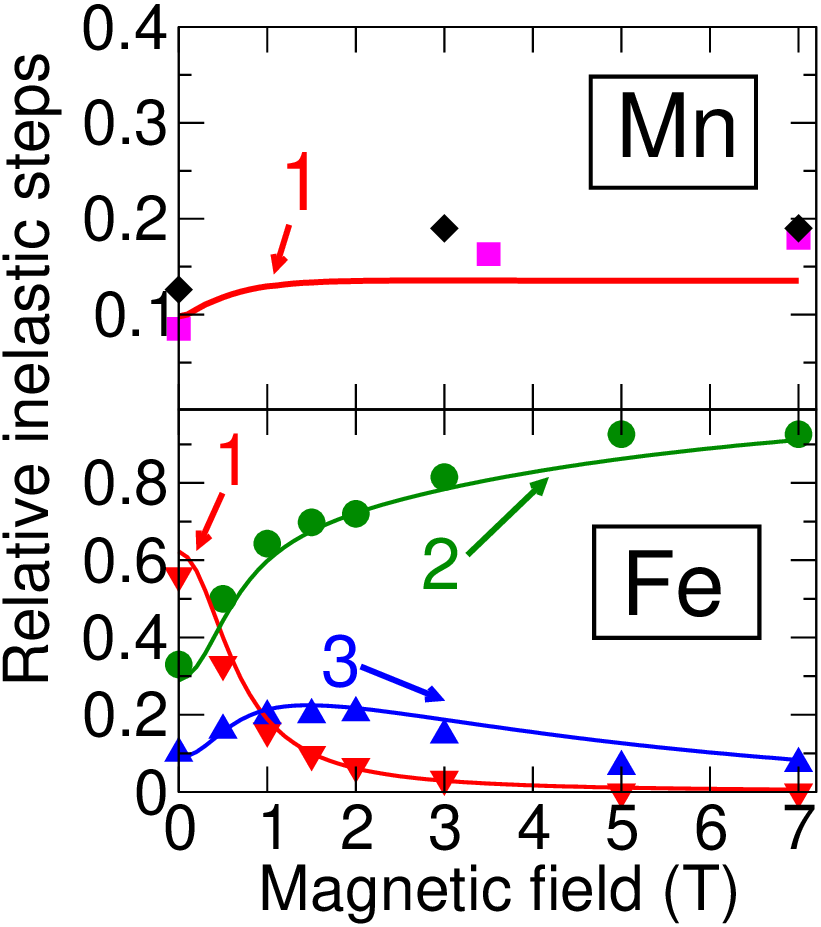}
\caption{(Color online)
Relative inelastic step heights in the conductance for Mn and Fe adsorbates 
as a function of the applied magnetic field, B, (along the N axis): 
calculations (full lines) and experiment~\cite{Hirjibehedin06,Hirjibehedin07} (symbols).
\label{figure3}}
\end{figure}

Fig.~\ref{figure2} presents the conductance as a function of the STM bias 
obtained as the product of the computed global conductance (Fig.~\ref{figure1}) by the 
 elastic/inelastic branching ratio
 from expression (\ref{fraction}). Results are shown for 
Fe adsorbates at five values of the B field (B along the N axis in 
part (a) and along the hollow axis in part (b)). A gaussian broadening of 
0.26 meV corresponding to a temperature of 0.5 K~\cite{LauhonHo} 
has been added. In this system, the Fe spin is equal to 2~\cite{Hirjibehedin07} so that the conductance 
can present 4 steps associated to the inelastic thresholds (labelled 1-4 on the figure). 
As a first remark, the contribution of inelastic channels is very large; 
for $B = 0$ and  for an infinite resolution in this system, the inelastic channels at large bias amount 
to around 67\%  of the elastic channel. At finite resolution, for B=0, the increase of the conductance 
between 0 and 8 mV  is smaller due to the small energy difference between $\phi_{0}$ and $\phi_{1}$.  
Second, Fig.~\ref{figure2} shows an important change in the inelasticity spectrum with B. The 0-1 excitation 
is dominating at low B and  disappears when B increases, whereas 0-2 dominates at large B. 
The 0-4 excitation is always weak. This behaviour is exactly the one observed experimentally~\cite{Hirjibehedin07}. 
For a quantitative comparison, Fig.~\ref{figure3} presents the relative step heights 
(ratio of the height of  a given  inelastic step  to the sum of the inelastic steps 1-3) 
as a function of B, compared with the experimental values. The 0-4 excitation is predicted to be very small and 
it is not observed experimentally for this geometry; we have not included it
 on the figure.  Results obtained for the other orientations of the B-field also reproduce the importance of inelastic channels.

Experiments on Mn adsorbates on CuN monolayers on Cu~\cite{Hirjibehedin07} showed a very small 
magnetic anisotropy  associated with a spin 5/2 and at finite resolution, the conductance is 
basically exhibiting a single inelastic step for all B values. Fig.~\ref{figure3} presents a 
comparison of our prediction for the relative inelastic step height (ratio of the inelastic step 
to the conductance at 0 bias) as a function of B, it is seen to be in quantitative agreement 
with the experimental data.

In the present approach, the excitation process is seen as a decoupling/recoupling process 
induced by the collision with the tunnelling electron. As our theory predicts,
spin excitation can take place without spin-flip of the electron flux.
We can distinguish two regimes as the magnetic field, B, increases. 
These are associated with a change in the magnetic structure of the system: 
 evolution from a magnetic anisotropy induced by the 
lattice at low B, towards  the Zeeman effect at higher B. 
The main axis of the lattice-induced anisotropy is the N axis (z-axis) and thus, 
the efficiency of the B field in generating a Zeeman structure is weaker along the 
hollow axis, as compared to the N axis (see Fig.~\ref{figure2}).
This change of the energy landscape has bearings on the actual
spin composition of the atom states.
As an example, the 0-1 excitation of Fe, see Fig.~\ref{figure3}, 
comes from the decoupling with the environment at low B, and  
at large B, the excitation is mainly due to coupling to the electron spin.
 Furthermore, in Fig.~\ref{figure3}, for incident electrons polarized along the N axis, 
the 0-1 and 0-4 transitions are not associated to a change of the collisional electron 
spin direction, whereas the 0-2 and 0-3 transitions are entirely spin-flip transitions.

In the case of Mn, the environment-induced anisotropy is very weak and 
for finite B, the Mn spin structure is a simple Zeeman splitting; 
in this case, the transitions are only spin-flip with a $\Delta M= \pm1$ selection rule 
and the fraction of inelastic tunnelling is basically given by a
ratio of squared Clebsch-Gordan 
coefficients leading to a nice agreement with experiment, 
see Fig.~\ref{figure3}.

The importance of the inelastic conductance in the total conductance 
is a direct consequence of the nature of the excitation process, 
analysed above. The initial magnetic state of the adsorbate couples 
with the spin of the collisional electron to form a collisional 
intermediate with a given total spin. At the end of the collision, 
the collisional intermediate populates all the possible asymptotic 
channels according to their weight in the intermediate, 
Eq.~(\ref{fraction}). 
The importance of a particular inelastic channel is then given by its weight 
in the collision intermediate (environment-induced anisotropy 
or spin coupling coefficient) and is not proportional to the 
modulus square of a matrix element
between initial and final states. 
This explains the large difference of inelasticity observed in 
magnetic IETS compared to vibrational IETS. 

However, inelastic rotational excitation shares many features with 
magnetic IETS.
The resonant rotational excitation of an adsorbed 
molecule induced by tunnelling electrons also involves 
 a transient angular momentum 
coupling between the collisional electron and the  molecule. 
The corresponding process can be formulated in a very similar way
to the present work and indeed leads to the observed strong rotational 
excitation~\cite{RotationO2,Stipe2}.

We believe that the present mechanism, which explains the 
strength of magnetic IETS and accounts for the observations 
in the case of Fe and Mn adsorbates, is of general occurrence.
In addition, the present formalism  yields a very easy way of accuratly
predicting the importance of spin transitions. One can stress that,
in the present approach, once the spin state of the collisional intermediate is fixed,
 all the spin changing transitions are fully determined. 
Indeed, the above formalism also leads to the quantitative account~\cite{nous} 
of the magnetic  IETS of Mn$_2$, Mn$_3$~\cite{Hirjibehedin06} 
and Co-phthalocyanine layers~\cite{Pc}. 
In the latter case, we find an extremely strong inelastic 
fraction in tunnelling, up to 300\% for 3 molecular layers, 
in excellent agreement with the experimental findings.

\begin{acknowledgments}
We thank Dr. Sebastian Loth for very interesting discussions.
We also thank
 Dr. Frederico D. Novaes for his help and important input.
Financial support from the Spanish MEC (No. FIS2006-12117-C04-01)
 is gratefully
acknowledeged.
\end{acknowledgments}


\end{document}